\documentclass[journal=cmatex,manuscript=article]{achemso}

\usepackage[version=3]{mhchem} 
\usepackage{placeins}
\usepackage{amsmath}
\usepackage{booktabs}       
\usepackage{amsfonts}       
\usepackage{nicefrac}       
\usepackage{microtype}      
\usepackage{xcolor}         
\usepackage{amsmath}
\usepackage{graphicx}
\usepackage{mathtools, amssymb, lipsum, nccmath} 
\usepackage{placeins}
\usepackage{enumitem} 
\usepackage{siunitx}
\usepackage{nameref}
\usepackage{lineno}

\newif\ifshowchanges

\author{Ashutosh Srivastava}
\affiliation[mse]
{School of Materials Science and Engineering, Georgia Institute of Technology, 771 Ferst Drive, Atlanta, GA 30332}
\author{Sakshi Agarwal}
\affiliation[mse]
{School of Materials Science and Engineering, Georgia Institute of Technology, 771 Ferst Drive, Atlanta, GA 30332}
\author{Shivank Shukla}
\affiliation[mse]
{School of Materials Science and Engineering, Georgia Institute of Technology, 771 Ferst Drive, Atlanta, GA 30332}
\author{Harikrishna Sahu}
\affiliation[mse]
{School of Materials Science and Engineering, Georgia Institute of Technology, 771 Ferst Drive, Atlanta, GA 30332}
\author{Rampi Ramprasad}
\email{ramprasad@gatech.edu}
\affiliation[mse]
{School of Materials Science and Engineering, Georgia Institute of Technology, 771 Ferst Drive, Atlanta, GA 30332}

\title[An \textsf{achemso} demo]
{On-the-Fly Machine-Learned Force Fields for High-Fidelity Polymer Glass Transition Simulations}

\begin{document}

\begin{abstract}
Predicting polymer glass transition temperatures ($T_g$) with first-principles fidelity has long remained out of reach, as cooling multi-thousand-atom systems over a broad temperature range at acceptable rates exceeds the computational limits of ab initio molecular dynamics (AIMD). Here we employ a hybrid scheme that merges AIMD with accelerated on-the-fly (OTF) machine-learned force-field (MLFF) construction, enabling $T_g$ prediction at quantum-mechanical accuracy with near-classical computational cost. The OTF protocol to construct MLFFs adaptively triggers first-principles calculations only when newly encountered configurations lie outside the current model’s domain of confidence, allowing robust, parameter-free MLFFs to be built from merely $\sim$1000 AIMD-sampled configurations per polymer. These MLFFs are then utilized to perform long-time cooling simulations on amorphous supercells containing several thousand atoms. Applied across twelve polymers spanning aromatic, aliphatic, heteroatomic, and branched chemistries, the method yields predictions in excellent accord with experiment while reducing computational cost by approximately six orders of magnitude relative to AIMD. This work establishes a new paradigm for predictive polymer modeling, demonstrating that OTF-MLFFs provide a generalizable, accurate, and scalable route to simulating the thermophysical behavior of complex disordered materials at near quantum-mechanical fidelity.

\end{abstract}

\section{Introduction}
High-fidelity simulations of complex macromolecular systems remain one of the central challenges in computational materials science. Many phenomena of practical and fundamental interest, including structural reorganizations in dense or disordered environments, thermophysical transitions, and collective fluctuations, unfold over nanoseconds or longer and involve thousands of atoms.\cite{Tran2024-int1,Muench2016-int2,Hager2015-int3,Huan2016-int4,Pan2021-int5} Polymers are emblematic of this challenge: their behavior is governed by slow configurational rearrangements, broad distributions of local environments, and pronounced dynamic heterogeneity that cannot be captured within the spatial or temporal windows accessible to conventional ab initio molecular dynamics (AIMD). Although AIMD provides unparalleled quantum-mechanical fidelity, its steep computational scaling confines simulations to small unit cells and/or short trajectories, rendering it inadequate for capturing the full physics of polymeric materials.

In recent years, machine-learned force fields (MLFFs) have emerged as a transformative pathway for accelerating atomistic simulations toward quantum accuracy. Frameworks based on pretrained or transferable architectures, including NequIP\cite{Batzner2022-a}, Allegro\cite{Musaelian2023-a}, MACE\cite{Kovcs2025-a}, and DeepMD\cite{Qi2025-a}, as well as kernel-based approaches, have significantly reduced the amount of system-specific training data required through transfer learning and fine-tuning strategies. Nevertheless, accurately describing polymeric systems remains challenging because their structural diversity, long-timescale dynamics, and complex phase behavior are often underrepresented in the datasets used to develop foundation models. As a result, additional polymer-specific training data and fine-tuning are frequently required to achieve reliable predictions. For example, recent benchmarking of MACE-OFF demonstrated limited transferability for several polymers, including Polyacrylonitrile (PAN), Polyethylene terephthalate (PET), and Poly[thio bis(4-phenyl)carbonate] (PTPC), while simulations of Polychlorotrifluoroethylene (PCTFE) were found to be unstable.\cite{Kovcs2025-a,arxivmicrosoft} Consequently, although foundation models substantially reduce the cost of force-field development, obtaining highly accurate MLFFs for polymer simulations can still require considerable data curation, model refinement, and computational resources.
These issues highlight the challenges in achieving both stability and physical fidelity in MLFF-driven polymer simulations. For polymers, where a complex hierarchy of primary and secondary interactions coexists and where configurational diversity is vast, such data demands and training burdens have impeded widespread application.

\begin{figure}[!ht]
    \centering
    \includegraphics[width=0.8\linewidth]{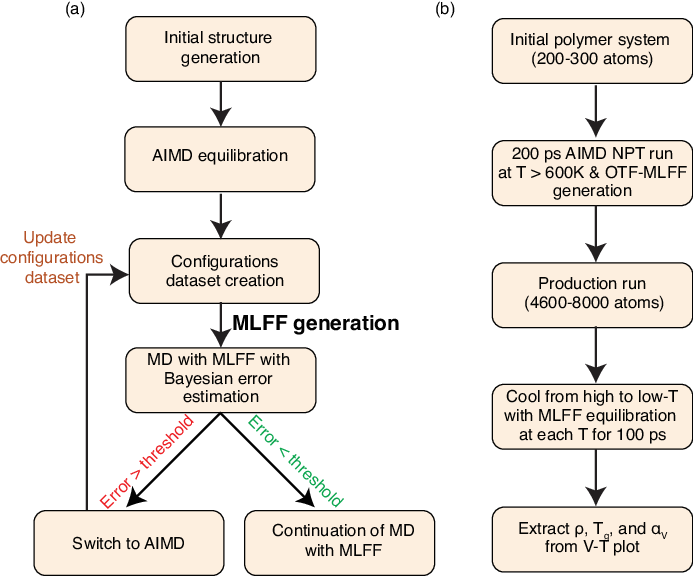}
    \caption{Adaptive on-the-fly machine-learning force-field (OTF-MLFF) framework for polymer thermophysical prediction. (a) Stepwise protocol for generating MLFFs, beginning with initial AIMD equilibration, Bayesian error estimation, automated dataset expansion, and real-time force-field refinement. (b) Workflow for predicting polymer glass transition temperature ($T_g$) and subsequent long-time ML-driven cooling trajectories enabling high-resolution volume–temperature (V-T) analysis to obtain density ($\rho$), $T_g$, and volume thermal expansion coefficient ($\alpha_V$).}
    \label{fig1}
\end{figure}

Hybrid QM/ML strategies offer an appealing alternative by invoking first-principles calculations only when needed. Among these, on-the-fly (OTF) MLFF learning stands out for its adaptivity, sample efficiency, and elimination of pre-generated training sets.\cite{Botu2014} In the OTF-MLFF paradigm, a simulation begins with a short AIMD segment that seeds an initial MLFF. As the trajectory proceeds, a Bayesian uncertainty estimator identifies configurations insufficiently represented in the model. Only in these cases is a new first-principles calculation triggered\cite{Jinnouchi2019-1}, after which the training set is expanded and the MLFF refined in real time (Figure~\ref{fig1}(a)). As configurational diversity increases during the run, the frequency of AIMD evaluations rapidly diminishes, and the force field becomes progressively more robust. The result is a parameter-free, ready-for-use, and physically grounded MLFF that approaches AIMD accuracy (within the domain of the original simulations) while operating near-classical computational cost.

Despite these advances, predicting the glass transition temperature ($T_g$) of polymers with first-principles fidelity has remained out of reach. $T_g$ marks a fundamental thermophysical transformation, a second-order transition separating rubbery and glassy regimes, and dictates mechanical performance, thermal stability, and processability.\cite{Roos2010, han1994glass, Park2024, yu2001polymer} Accurate $T_g$ estimation requires slow cooling from the melt, precise detection of subtle volumetric discontinuities, and ample sampling of heterogeneous dynamics, conditions that challenge the computational capabilities of AIMD\cite{han1994glass, yu2001polymer, buchholz2002cooling, mohammadi2017glass}. Classical molecular dynamics can perform long cooling trajectories but relies on empirical force fields whose transferability across polymer families, chemistries, and architectures may be unreliable or is sometimes infeasible. Recent strides in simulating the glass transition phenomenon in polymers using standalone MLFFs such as Vivace\cite{arxivmicrosoft} and MACE-OFF\cite{Kovcs2025-a}, while promising, also highlight key challenges- likely rooted in the careful selection of the training data. These approaches either fail to reveal the glass transition with sufficient clarity or fail entirely for certain polymer chemistries (see Figures S7 and S8 in the supplementary material of the work\cite{arxivmicrosoft}).

Here, we present the first general and scalable OTF-MLFF\cite{Jinnouchi2019-1, vandermause2020fly} framework capable of simulating polymer $T_g$ with ab initio fidelity across a broad chemical space, enabled by the capability implemented in the Vienna Ab-initio Simulation Package (VASP)\cite{kresse1996efficient, kresse1996efficiency}.
We evaluate twelve polymers spanning aromatic, aliphatic, heteroatomic, and branched architectures. For each system, a MLFF is generated from a 200-ps NPT equilibration at elevated temperatures (600–800 K), using only $\sim$1000 quantum-sampled configurations. These MLFFs are then deployed to equilibrate large amorphous supercells containing several thousand atoms and to perform gradual cooling from the melt to cryogenic temperatures (Figure~\ref{fig1}b). From the resulting volume–temperature profiles, $T_g$, density, and thermal expansion coefficients are extracted. 
Across all polymers, the predicted quantities exhibit excellent agreement with experiment while reducing computational cost by roughly six orders of magnitude relative to a full AIMD treatment.

More broadly, this work establishes a new paradigm for physics-driven predictive polymer modeling, delivering near-quantum-mechanical accuracy at near-classical computational expense. By making high-fidelity, large-scale simulations of polymers tractable on modest computational resources, this approach unlocks unprecedented opportunities to interrogate complex polymer phenomena. 

\section{Computational Methodology}
\textbf{Structure Generation:}
All the initial polymer structures were generated using an updated Polymer Structure Predictor (PSP)\cite{Sahu2022PSP}. As the RDKit\cite{rdkit_zenodo} often fails to generate three-dimensional (3D) structures for large or highly branched polymers, its distance–geometry–based embedding algorithm struggles to satisfy competing spatial constraints, resulting in convergence issues and steric clashes during structure generation. To address these challenges, PSP was enhanced to first construct and optimize 3D geometries of monomer and end-cap units individually, followed by their iterative assembly into polymer chains with periodic relaxation using the universal force field (UFF), thereby yielding robust and physically realistic initial geometries without embedding failures. For the construction of amorphous polymer models, PSP employs PACKMOL\cite{Martnez2009pkmol} to pack polymer chains within a simulation box while allowing rigid-body rotations and translations; however, this procedure often results in systems with densities lower than the target value. To overcome this limitation, PSP now incorporates an equilibration stage in LAMMPS\cite{plimpton1995fast} using the GAFF2 force field and the 21-step relaxation protocol proposed by Abbott et al.,\cite{Abbott2013-eq21step}, enabling efficient structural relaxation and equilibration of the amorphous models at the desired temperature.

\textbf{Molecular Dynamics Details:}
The generated structures were then equilibrated, and a force-field was simultaneously trained using the OTF approach\cite{Jinnouchi2019-1, Jinnouchi2019-2, Jinnouchi2020} as implemented in VASP (version 6.5.1)\cite{kresse1996efficient, kresse1996efficiency}. The OTF learning approach is a hybrid approach of first principles and machine learning. First-principles calculations were only executed when a new configuration was encountered for which the Bayesian error exceeds the threshold and which was not sampled earlier during the ab-initio molecular dynamics (AIMD) run.\cite{Jinnouchi2019-1} Similarly, for the configurations that were very close to the configurations already sampled, the approach uses a machine learning algorithm to predict the first-principles output parameters such as forces, energies, and stresses. In this way, we collect all the distinct configurations along with the atomic positions and forces, which are then utilized to build the machine learning force field. OTF-AIMD simulations have been performed using an isothermal–isobaric (NPT) ensemble, allowing the cell shape and volume to relax during dynamics. The Langevin thermostat\cite{Hoover1982,Evans1983} is employed to maintain temperature and pressure, utilizing the Parinello-Rahman algorithm\cite{Parrinello1980,Parrinello1981}. 
The training simulations have been performed for 200 ps (500,000 steps) with an optimized step size of 0.4 femtoseconds for polymeric structures containing lighter elements. To facilitate reasonable phase-space sampling, the initial temperature was maintained at 600 K (800 K for polymers with $T_g$$\ge$400 K), and the final temperature was set 30\% higher, 780 K (1040 K). The pressure was kept at 1 atm during the run.

\textbf{Electronic Structure Calculation Details:}
The plane-wave cut-off energy for the electronic charge density expression was set to 500 eV, and the overall precision of the numerical integration was set to \textit{Accurate} to obtain reliable forces and stresses during the run. Inter-chain van-der Waals interactions were captured using the DFT-D3 scheme\cite{grimme2010consistent} with Becke–Johnson damping\cite{Grimme2011}. The partial electronic occupancies were treated using Fermi-smearing with a smearing width of 0.1 eV. The exchange-correlation was treated by Projector-augmented wave (PAW) pseudopotentials\cite{blochl1994projector,kresse1999ultrasoft} under the generalized gradient approximation (GGA). A 1$\times$1$\times$1 $\Gamma$-centered $k$-mesh was used to sample the Brillouin zone. The electronic minimization was performed using a mixture of the blocked-Davidson and Residual Minimization Method with Direct Inversion in the Iterative Subspace (RMM-DIIS) algorithms,\cite{Kresse1996-01,Kresse1996-02,Pulay1980} with a convergence criterion of 0.00001 eV during the self-consistent loop. 
 
\textbf{High-to-Low Temperature Cooling Details:}
The constructed MLFFs for each polymer were subsequently applied to the corresponding larger polymer structures, each having $\sim$20 to 32 polymer chains (see SI-Table 2 for details). Each system was further equilibrated using the respective MLFFs, starting from the training temperature and cooling down to 50 K in 25 K intervals. At each temperature step, simulations were run for 100 ps, and the final 50 ps of each trajectory was used to construct the volume–temperature (VT) curve.
To optimize storage efficiency without compromising accuracy, volume data were recorded every ten steps and used to calculate the average and standard deviation at each temperature. The resulting VT curves were fitted with two linear segments, and the intersection point of these lines was taken as the $T_g$ of the corresponding polymer. Additionally, an error-propagation analysis was performed to estimate the 66.67\% confidence interval (CI) of $T_g$, based on the uncertainties in the fitted slopes (refer to the Supplementary Information for details). 

\section{Results and Discussion}
\textbf{Establishing the approach using polyethylene:}
\begin{figure}[!h]
    \centering
    \includegraphics[width=1.0\linewidth]{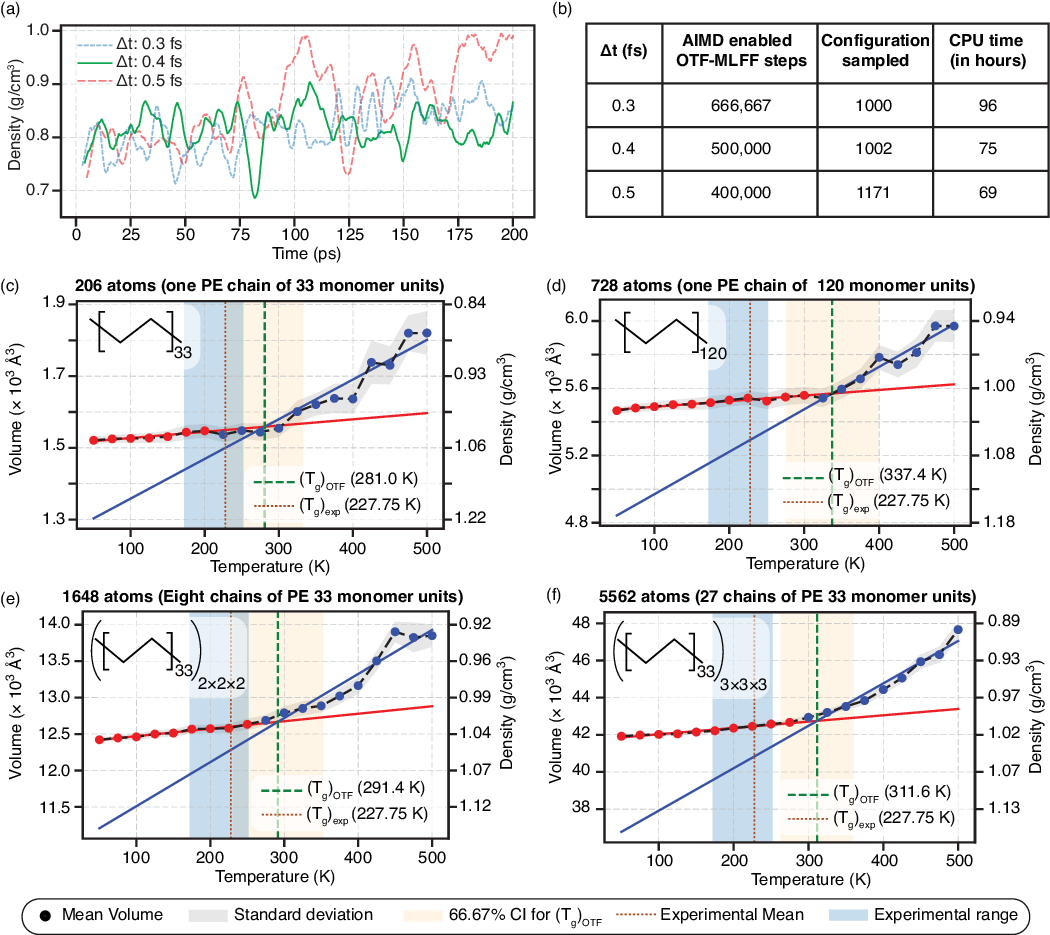}
    \caption{Thermodynamic behavior and computational performance of machine-learned force-field simulations across system sizes. (a) Density as a function of simulation time for different integration step sizes ($\Delta$t = 0.3, 0.4, and 0.5 fs) at 600 K. (b) Summary of timestep selection, number of AIMD-enabled OTF-MLFF steps required for a 200 ps trajectory, configurations sampled, and total CPU time (hours). (c–f) Volume (density)–temperature relations for polyethylene systems containing 206, 728, 1648, and 5562 atoms constructed from multiple chains of varying lengths.}
    \label{fig2}
\end{figure}
In AIMD simulations, the integration time step ($\Delta t$) is a critical parameter that controls both accuracy and numerical stability. Larger time steps reduce computational cost but risk significant errors in energies and forces, while smaller time steps improve fidelity at the expense of substantially higher computational effort. This balance is especially important for polymers, where high-frequency vibrational modes—dominated by C–H motions—impose stringent requirements for stability.
To assess the impact of $\Delta t$, we developed a $T_g$ simulation protocol using polyethylene (PE), chosen for its simplicity and well-characterized behavior. An amorphous single-chain 33-mer PE structure (composed of 206 atoms) was generated at 300 K using the Polymer Structure Predictor (PSP)\cite{Sahu2022PSP} and subsequently equilibrated at 600 K for 200 ps using AIMD with OTF-MLFF. Density profiles were extracted for $\Delta t$ values of 0.3, 0.4, and 0.5 fs. As shown in Figure \ref{fig2}(a). Detailed information on $\Delta t$, number of AIMD-assisted OTF-MLFF steps, configuration sampled, and CPU time has been mentioned in Figure~\ref{fig2}(b). The 0.5 fs simulation exhibited large density fluctuations, indicating numerical instability, whereas the 0.3 and 0.4 fs simulations produced nearly identical, well-behaved density trajectories. Both yielded comparable sampling quality (1000 vs. 1002 configurations) used to create MLFF, but with a notable difference in cost: the 0.3 fs simulation required $\sim$30\% more wall time (96 h) than the 0.4 fs run (75 h).
Balancing accuracy and efficiency, a time step of $\Delta t = 0.4$ fs was identified as the optimal choice and was therefore used for all subsequent simulations in this study.

The local atomic environments sampled during the high-temperature equilibration were used to construct a production-ready MLFF. This MLFF was then used to calculate the $T_g$ of the 33-mer amorphous PE structure obtained at the end of the 600 K AIMD with OTF-MLFF simulation. The system was cooled from 600 K to 50 K in 25 K decrements, equilibrating for 100 ps at each temperature. The final configuration from each step served as the starting point for the next, and the average volume was computed from the last 50 ps of each trajectory segment to ensure thermal equilibration.
The resulting volume–temperature curve (Figure~\ref{fig2}(c)) was fitted to extract the glass transition temperature, with the $T_g$ value indicated by the dotted vertical green line. The emergence of two distinct linear regimes in the volume–temperature relationship further demonstrates the model’s ability to faithfully reproduce the second-order phase transition behavior characteristic of polymer glass transitions.
A 66.67\% confidence interval, obtained via error-propagation analysis (Supplementary Information), is shown as the orange-shaded region, while the gray-shaded regions represent the standard deviation of the volume at each temperature. The computed $T_g$ of 281 ± 51.7 K is in reasonable agreement with the experimental range reported for a variety of polyethylene flavors\cite{fakirov2000pe1, Ohlberg1958pe2}, indicated by the blue-shaded region.
Fluctuations in the average volume at elevated temperatures are primarily attributed to finite-size effects inherent to the 33-mer model. Nonetheless, the MLFF exhibits strong stability and transferability across the full temperature range. 

\textbf{Testing scalability across system sizes:}
Next, we consider PE systems with longer chain lengths and significantly larger simulation cells containing multiple repeat units. The monomer definitions and structural details of the various amorphous PE models are provided in SI-Table 1, and the corresponding $T_g$-derived volume–temperature curves are shown in Figures~\ref{fig2}(c–f). As the system size increases, the volume/density fluctuations during cooling diminish noticeably, yielding smoother and more stable volume/density–temperature profiles. This effect is particularly apparent in the 5562-atom system (Figure~\ref{fig2}(f)), where local structural variations are averaged out more effectively, leading to a more uniform macroscopic response. Moreover, as shown in Figure 2(c-f), the densities across different PE system sizes remain consistent, strengthening the MLFF transferability claim.

These results demonstrate that an MLFF trained on a relatively small configuration set remains robust when transferred to substantially larger systems of the same chemical composition. The model preserves accuracy and stability across both system size and temperature, underscoring the inherent scalability of the OTF-MLFF framework. Furthermore, from the $T_g$ simulations, we extracted the coefficient of thermal expansion (CTE) in the glassy and rubbery regimes. Across all PE systems studied, the CTE ranges from 60.5$\times$10$^{-6}$ - 113.4$\times$10$^{-6}$ K$^{-1}$ below $T_g$ and 438.6$\times$10$^{-6}$ – 671.2$\times$10$^{-6}$ K$^{-1}$ above $T_g$, in good agreement with the reported experimental values in the range $\sim$93$\times$10$^{-6}$ K$^{-1}$ - 600$\times$10$^{-6}$ K$^{-1}$\cite{CLTE1,CLTE2,Ohlberg1958pe2} (for details refer to SI-Table 1). This confirms that the MLFF not only reproduces $T_g$ behavior but also captures the fundamental thermophysical physics associated with polymer relaxation and thermal expansion. 

We performed an independent replica $T_g$ simulation for the 5562-atom PE system using the MLFF trained on the 206-atom dataset to further assess the robustness and transferability of the developed potential. The equilibration protocol employed in the replica simulation was identical to that used in the original calculation (see Figure~\ref{fig2}(f)), ensuring a consistent basis for comparison. The resulting temperature-dependent volume/density profiles, along with their corresponding standard deviations, are presented in SI-Figure 1(a). The close agreement among the volume evolution, density trends, and fitted $T_g$ values across the independent simulations demonstrates excellent reproducibility of the predicted glass transition behavior. Furthermore, the relatively small variations observed between the replica runs indicate that the calculated $T_g$ is not sensitive to the choice of initial configuration or stochastic fluctuations during equilibration. These results provide strong evidence that the MLFF trained using a relatively small 206-atom reference system can reliably capture the thermodynamic behavior of substantially larger PE systems. Overall, the consistency of the replica simulations confirms the robustness, stability, and transferability of the developed MLFF for predicting glass-transition-related properties in PE.

The cooling rate is a critical parameter in $T_g$ simulations because glass transition is inherently a kinetic phenomenon. It defines the rate at which the temperature of a material decreases with time and directly influences the extent of molecular relaxation during cooling. For a given polymer, a higher cooling rate provides less time for the polymer chains to relax toward equilibrium, generally resulting in a higher observed $T_g$. Consequently, experimental measurements typically employ relatively slow cooling rates (approximately 5--20 K/min) to allow sufficient molecular relaxation and structural rearrangement.\cite{gaisford2016principles,menczel2009thermal}

In contrast, atomistic molecular dynamics simulations are restricted to much higher cooling rates, typically in the range of 10$^{11}$-10$^{13}$ K/min, owing to computational limitations.\cite{Wang2020} Despite this discrepancy of nearly 10$^{12}$ K/min compared to experiments, previous studies have shown that simulated $T_g$ values obtained at such cooling rates remain in reasonable quantitative agreement with experimental measurements.\cite{Wang2020,Soldera2006} To evaluate the effect of cooling rate within the present OTF-MLFF framework, we performed $T_g$ simulations for PE using two different cooling rates, 1.5$\times$10$^{13}$ K/min and 7.5$\times$10$^{12}$ K/min, comparable to those employed by Soldera \textit{et al.}\cite{Soldera2006} The resulting $T_g$ values were nearly identical, as shown in SI-Figure 1(b), indicating that either cooling rate can be reliably used in subsequent OTF-MLFF simulations.

In principle, the determination of polymer $T_g$ should be performed in the infinite-chain-length limit. However, atomistic simulations are restricted to finite chain lengths because of computational cost. To assess the adequacy of the chosen system size, we conducted a series of $T_g$ simulations for PE using the developed MLFF, considering 27 polymer chains with chain lengths ranging from 5 to 40 monomer units (5-, 10-, 15-, 20-, 25-, 30-, and 40-mers). The 35-mer system was omitted because results were already available for a comparable 33-mer system.
The simulated $T_g$ values were analyzed using the well-known Flory--Fox relation:\cite{Fox1950}
\begin{equation}
    T_g = T_g^{\infty} - \frac{K}{M_n}
    \label{eq:1}
\end{equation}
where $T_g^{\infty}$ is the glass transition temperature at infinite molecular weight, (K) is an empirical constant, and ($M_n$) is the number-average molecular weight. Extrapolation of the simulation data using Eq.~\ref{eq:1} yielded an infinite-chain-length glass transition temperature of 340.48 $\pm$ 16.67 K (see SI-Figure 2
and Supporting Information for details). This value is in close agreement with the $T_g$ obtained for the system containing 27 chains and 5562 atoms (Figure~\ref{fig2}(f)). The agreement suggests that a simulation cell containing approximately 5000 atoms is sufficient to capture the glass transition behavior of PE within the OTF-MLFF framework, while maintaining computational efficiency.

\textbf{Validating approach over different chemistries:}
To evaluate the generality of the OTF-MLFF framework across diverse chemistries, we selected a set of polymers spanning a wide range of structural complexity and experimental $T_g$ values. The SMILES representations\cite{Weininger1988-smiles} and molecular structures of all polymers are provided in SI-Table 2. For each system, an initial amorphous cell containing roughly 200–250 atoms was constructed using PSP; the corresponding number of monomer units and total atoms per cell are also summarized in SI-Table 2. This training system size reflects the practical computational limits of AIMD (based on the Figure~\ref{fig2}(b) analysis), ensuring that each MLFF could be generated within a feasible wall time. 
\begin{figure}[!ht]
    \centering
    \includegraphics[width=1.0\linewidth]{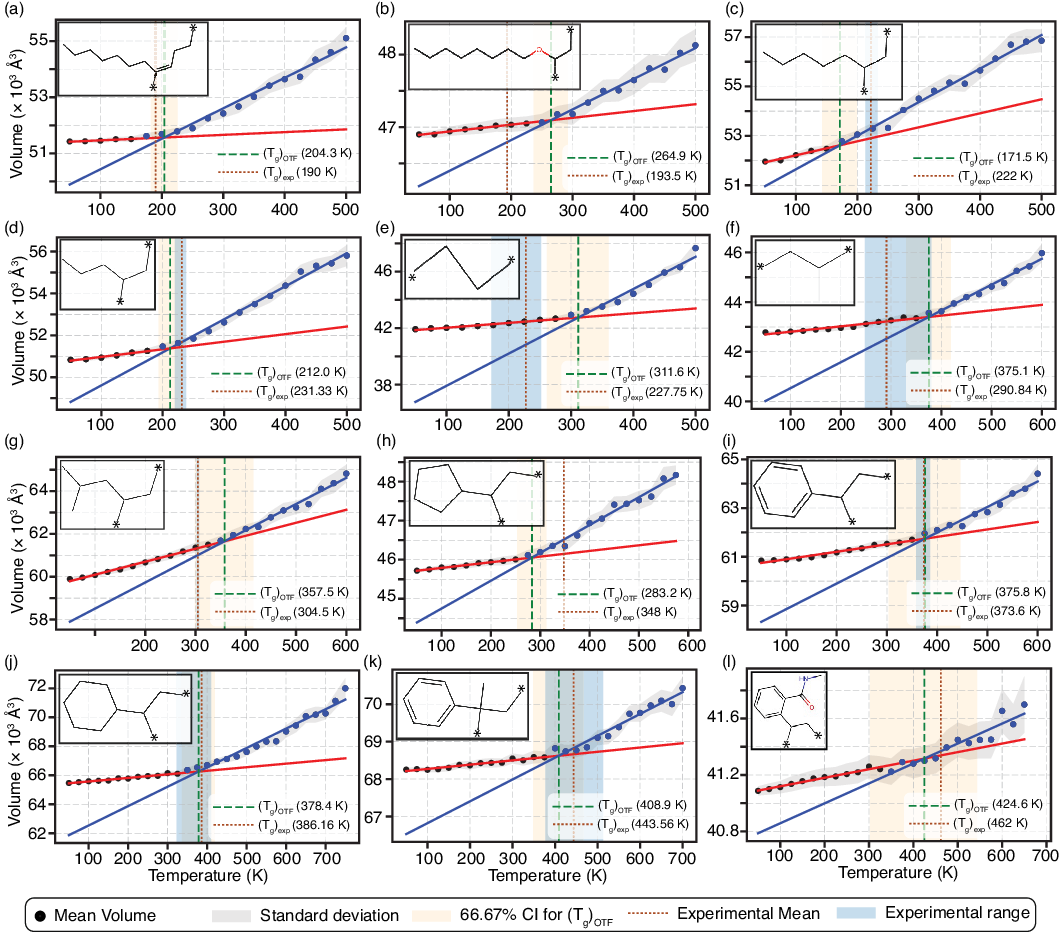}
    \caption{Volume–temperature behavior of the polymers considered in this study. The corresponding polymer structures are shown in each panel. Red and blue markers with shaded gray regions denote the average volumes and standard deviations obtained from the final 50 ps of the production trajectory. Dashed green lines with shaded orange bands indicate the calculated glass-transition temperatures ($T_g$) and their 66.67\% confidence intervals. Experimental average $T_g$ values are indicated by red dotted lines, and the shaded blue region indicates the experimental range of the reported $T_g$ values for comparison.}
    \label{fig3}
\end{figure}
Each polymer was equilibrated for 200 ps at a designated training temperature ($T_{training}$), chosen to sample a broad and representative range of atomic configurations suitable for subsequent $T_g$ simulations. The number of configurations collected during training ($N_{sampled}$) varied by chemical complexity and is listed in SI-Table 2. Leveraging the insights drawn from the PE case, we then considered significantly large structures for temperature-ramped simulations, typically 4600-8000 atoms constructed by replicating and combining multiple chains of each polymer.

The resulting volume–temperature curves, shown in Figures~\ref{fig3}(a)–(l), include the polymer structures, the standard deviation in equilibrated volumes (gray-shaded regions), and the fitted $T_g$ values with their associated confidence intervals (vertical dashed green lines). Corresponding density–temperature plots for all cases are provided in SI-Figure 3. Across all twelve polymers studied, the predicted $T_g$ values agree closely with the experimental values and lie within or close to the respective confidence bounds. This strong correlation underscores the predictive reliability of the OTF-MLFF approach, which successfully captures the glass transition behavior of polymers with aromatic, aliphatic, branched, and heteroatom-containing backbones.
Because the entire workflow is anchored in first-principles calculations and involves no empirical fitting or prior knowledge of the experimental $T_g$, the method is inherently transferable and broadly applicable. It provides a robust and universal pathway for simulating thermophysical properties of chemically diverse polymer systems well beyond the reach of traditional force-field approaches.

\begin{figure}[!ht]
    \centering
    \includegraphics[width=1.0\linewidth]{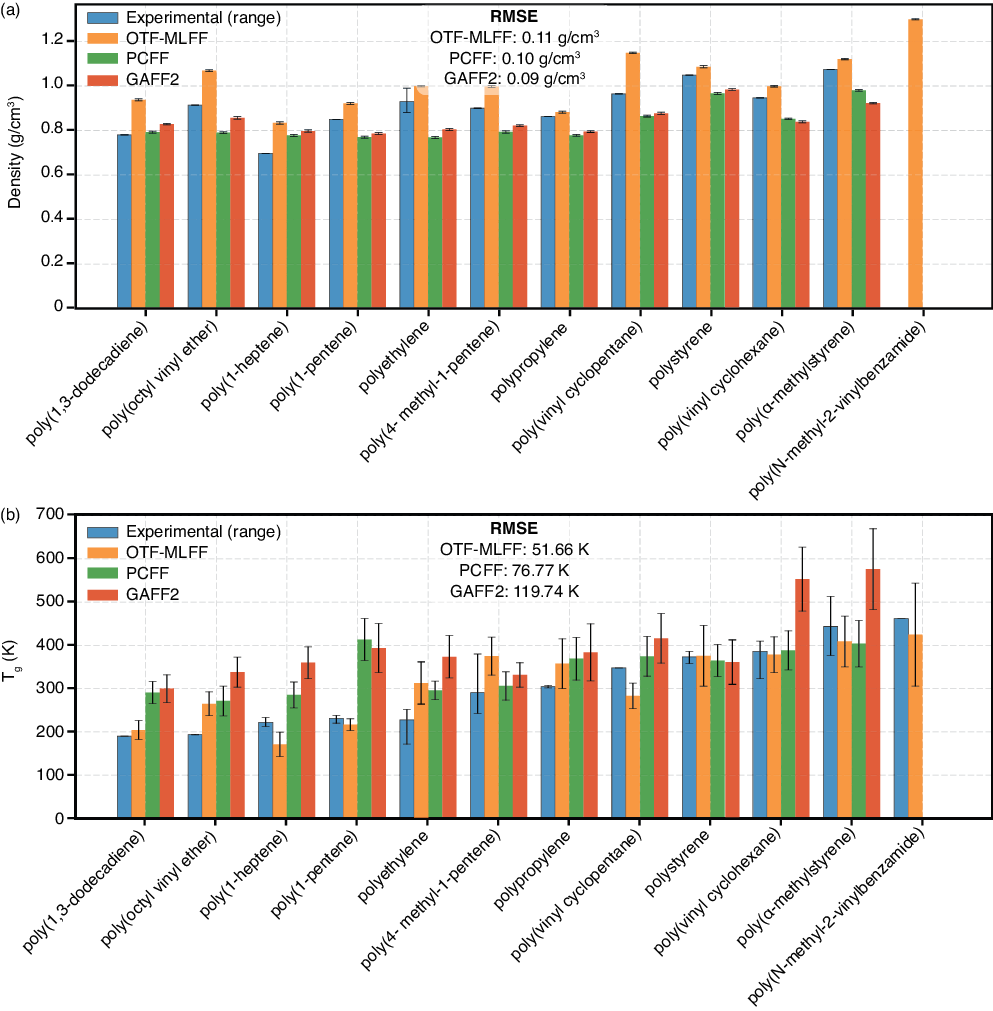}
    \caption{Comparison of OTF-MLFF approach predictions with reference data. (a) Densities at 300 K and (b) glass-transition temperatures ($T_g$) for the polymers investigated, benchmarked against experimental measurements and classical force-field results (PCFF and GAFF2). For the last polymer (poly(N-methyl-2-vinylbenzamide)), PCFF and GAFF2 simulations were not possible due to the unavailability of classical force-field parameters.}
    \label{fig5}
\end{figure}

The physical and chemical fidelity of the developed MLFFs is further confirmed by the excellent agreement between the predicted and experimental densities and glass transition temperatures across all polymers studied. For representative systems such as PE and polystyrene (PS), the experimental densities at 300 K of 0.88–0.93 and 1.05 g/cm$^{3}$, respectively\cite{Cheremisinoff2001, Gehlsen1995} come close to the MLFF estimates of 0.99 and 1.09 g/cm$^{3}$, respectively. Similar levels of consistency are observed for the remaining polymers, as illustrated in Figures \ref{fig5}(a) and (b), which compare the experimentally measured and MLFF-predicted $T_g$ values and 300 K densities. Further, the thermal expansion information was extracted from the volume-temperature plots (Figure~\ref{fig3}) of the respective polymers and is reported in SI-Table 3. These results highlight that the OTF-MLFF workflow effectively captures the essential chemical interactions and thermophysical behavior even in systems with complex chemistries, long chain lengths, and multiple repeat units, thereby reinforcing its scalability, robustness, and versatility. 

To further benchmark performance, we compared OTF-MLFF predictions with classical force-field simulations using PCFF and GAFF2. The RMSE values for density relative to experiment at room temperature are comparable across methods—0.11 g/cm$^{3}$ (OTF-MLFF), 0.10 g/cm$^{3}$ (PCFF), and 0.09 g/cm$^{3}$ (GAFF2), indicating that all three approaches reproduce equilibrium densities reasonably well at room temperature (Figure \ref{fig5}). In contrast, the advantage of OTF-MLFF becomes pronounced for glass-transition prediction, which is a straight test of performance over a broad temperature range: the RMSE for $T_g$ is significantly lower at 51.66 K for OTF-MLFF, compared to 76.77 K for PCFF and 119.74 K for GAFF2. The OTF-MLFF performance also surpasses refined OPLS\cite{Afzal2020refinedOPLS} and OPLS3e\cite{Roos2019opls3e} force fields, which report $T_g$ RMSE values of ~98.5 K\cite{Afzal2020refinedOPLS}.

Importantly, classical force-field simulations could not be performed for poly(N-methyl-2-vinylbenzamide) due to missing parameters, whereas OTF-MLFF successfully handled this system without modification of the simulation protocol, reflecting its intrinsic ab initio-derived generality. Collectively, these comparisons underscore the ability of OTF-MLFFs to accurately capture polymer thermophysical properties even for chemistries that lie beyond the domain of traditional force fields or even stand-alone MLFFs.
A useful aspect of our approach is that the second-order phase transition can be identified directly from the volume–temperature curves, without relying solely on mathematical fitting. This offers an alternative perspective to the recently introduced Vivace MLFF method \cite{arxivmicrosoft}, particularly in cases where the transition is difficult to detect. In their work, four of the ten polymers studied (PODPM, PVMS, PAN, and PTPC) exhibit nearly linear volume–temperature behavior in the entire temperature range considered with no discernible change in slope, suggesting that a glass transition may not be present according to their simulations. From a physical standpoint, the absence of any perceptible change in thermal expansion would indicate that fitting for an expected transition may not yeild in reilable transition temperature. It is a bit surprising that Vivace approach does not produce a discernible transition, even though they bracket the simulated temperature range around experimental $T_g$ value for the system studied.  
In our workflow, careful sampling from diverse regions of phase space during MLFF training facilitates the unambiguous detection of glassy and rubbery regimes when such transitions exist, even though an unbiased range of temperatures is used to capture the glass transition. This allows the method to avoid assigning $T_g$ where no transition is apparent and to provide more reliable estimates when a transition is observable.
 
\begin{figure}[!ht]
    \centering
    \includegraphics[width=1.0\linewidth]{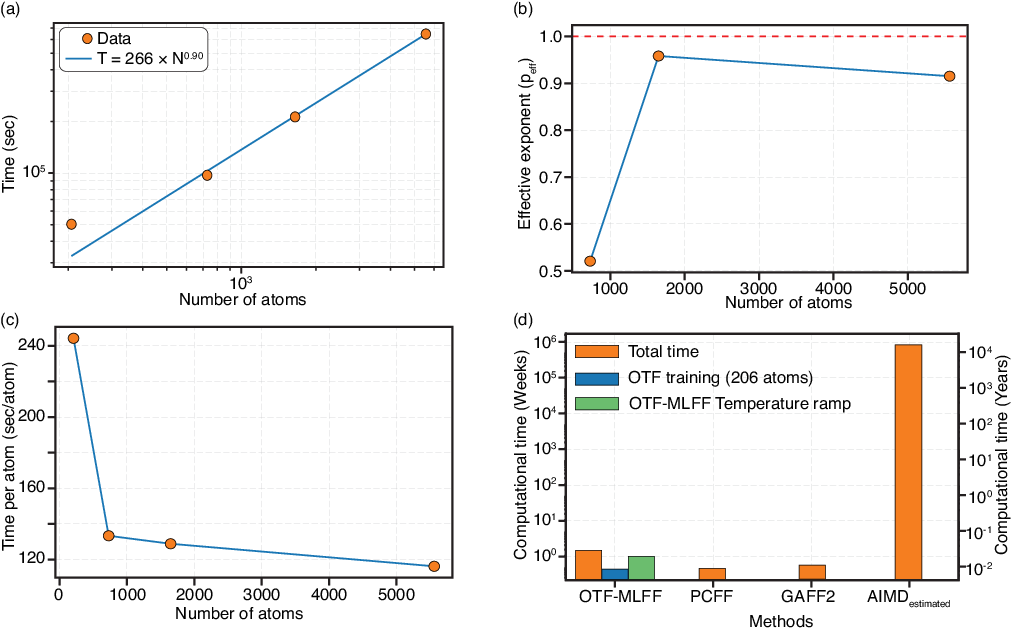}
    \caption{Scaling behaviour and computational benchmarking of the simulations. (a) log-log plot of simulation time with system size ($N$), (b) effective exponent ($p_{eff}$) as a function of $N$, (c) simulation time per atom versus $N$, and (d) Comparison of total simulation times for OTF-MLFF, the Polymer Consistent Force Field (PCFF), the General AMBER Force Field (GAFF2), and ab initio molecular dynamics (AIMD).}
    \label{fig4}
\end{figure}

\textbf{Computational time advantage:}
A notable outcome of this work is the ability of MLFFs trained on only $\sim$1000 AIMD configurations per polymer chemistry from 200–250-atom cells to remain accurate when applied to much larger systems. Using PE as a representative example, we now examined how the computational cost evolves with system size (Figure \ref{fig4}(a)). These MLFFs exhibit a near-linear (almost sublinear) scaling of $N^{0.90}$ ($N$ is the total number of atoms), in sharp contrast to the $N^{2}$-$N^{3}$ scaling of conventional DFT. This behavior highlights the practical scalability of the approach and its suitability for simulations of polymer structures containing several thousand atoms.

In order to confirm the (sub)linear scaling of the approach, the effective exponent ($p_{eff}$) and time per atom ($T/N$) as a function of number of atoms in the simulation ($N$) were calculated.
$p_{eff}$ is extracted from two consecutive points for the two neighboring system sizes $N_i$ and $N_{i+1}$ with corresponding computational times $T_i$ and $T_{i+1}$ using the following equation: 
\begin{equation}
    p_{eff}(N_i) = \frac{log(T_{i+1}/T_i)}{log(N_{i+1}/N_i)}
\end{equation}

Figure~\ref{fig4}(b) further examines the effective scaling exponent ($p_{eff}$) across system sizes. In all cases, $p_{eff}$ remains below or close to unity, confirming sub(linear) behavior. The initial rise, followed by a gradual potential saturation close to unity, reflects the crossover from small-system overheads to the regime where the MLFF achieves its characteristic efficient scaling. A complementary assessment using the simulation time per atom (Figure~\ref{fig4}(c)) similarly shows a monotonic decrease with increasing $N$. This behavior underscores both the scalability and the cost-effectiveness of the method while preserving \textit{ab initio} accuracy for large-scale polymer simulations.

The efficiency of the approach becomes particularly evident when comparing the computational cost of $T_g$	
simulations across methods. For polyethylene, training the OTF-MLFF using a 206-atom system for 200 ps and subsequently applying it to equilibrate a 5562-atom supercell for 100 ps at 23 temperature points (from 600 K to 50 K in 25 K intervals) amounts to roughly 6.25 million simulation time-steps. Using the OTF-MLFF framework, these steps were completed in approximately 10 days on 32 cores of an \textit{Intel(R) Xeon(R) Gold 6140 CPU @ 2.30 GHz}. The distribution of wall time between MLFF training and the $T_g$ cooling trajectory is shown in Figure~\ref{fig4}(d).

In stark contrast, performing the same number of simulation steps for the 5562-atom PE structure using AIMD would require an estimated $\sim$15,000 years on the same hardware. This estimate was determined by simply multiplying the time taken for a single time step, in the same hardware for the same system size, by 6.25 million.
Given that $T_g$ simulations inherently rely on serial cooling trajectories, such AIMD-based calculations become computationally prohibitive and likely cannot be practically scaled even on modern high-performance computing architectures. With the hybrid OTF-MLFF approach, we achieve an extraordinary speedup of approximately 10$^6$, while preserving ab initio fidelity in energies, forces, and stresses. Importantly, this acceleration does not compromise the predictive reliability of the model, underscoring the real-world applicability of OTF-MLFF for large-scale polymer simulations and the accurate prediction of complex thermophysical properties.

\textbf{Opportunities:}
While the OTF-MLFF framework is highly promising, several avenues remain for further refinement. One important direction is enhancing force-field robustness by integrating training data sampled across a broader temperature range, which could further improve transferability and performance in regimes with significant structural or dynamical variation. A temperature-diverse training strategy may reduce volume fluctuations during cooling trajectories, thereby improving the precision of $T_g$ estimation. However, achieving this within the current VASP-based OTF workflow is computationally demanding, and the benefits relative to the required cost are not yet fully established. This challenge highlights a deeper need for algorithmic and code-level optimization, both in the underlying AIMD engine and in the MLFF construction pipeline, to reduce overheads and enable more extensive sampling. Another direction and challenge is to develop a transferable MLFF that can be created by combining previously generated OTF-MLFFs without performing new simulations. Because a model for each polymer is trained independently, differences in descriptor distributions, energy baselines, and configuration-space coverage complicate direct integration. Several computational strategies can address this limitation, including descriptor harmonization through normalization or alignment, post-hoc energy, force, and stress re-referencing to ensure consistency, and data-fusion approaches that merge and prune training sets to produce a chemically diverse yet efficient reference database. Alternatively, modular or mixture-of-experts architectures could blend multiple MLFFs based on local environments, enabling smooth interpolation across polymer chemistries. Collectively, these approaches offer a pathway toward constructing a broadly transferable, low-cost MLFF by reusing existing OTF-generated data without repeated AIMD sampling. 

Looking ahead, the efficiency and near-first-principles accuracy of OTF-MLFFs open the door to entirely new classes of polymer simulations that were previously inaccessible. These include long-term thermal protocols, such as repeated cooling–heating cycles for accurate $T_g$ prediction, as well as mechanical deformation and fracture simulations that capture bond breaking and strain localization. Additionally, molecular and ionic transport studies may be conceived that involve the diffusion of gases, solvents, and ions through dense or heterogeneous polymer matrices. OTF-MLFFs may also enable simulations of degradation mechanisms and chemical transformations. By supporting these computationally demanding studies at scale, while maintaining quantum-level fidelity, OTF-MLFFs have the potential to significantly broaden the scope of polymer modeling and provide quantitative, mechanistic insights into phenomena that have long been beyond reach.

\section{Conclusion}
In conclusion, we present an on-the-fly machine-learned force field (OTF-MLFF) protocol for predicting the glass transition temperature ($T_g$) of polymers from atomistic simulations. By combining first-principles calculations, active learning, and MLFFs, the approach enables efficient simulations at scales that are challenging for direct ab initio methods while requiring only a limited number of training configurations. Across the polymers examined in this study, the resulting force fields reproduced physically reasonable density–temperature behavior and enabled direct estimation of $T_g$ from continuous cooling simulations without initialization near experimental values. These results demonstrate the potential of OTF-MLFFs as a practical and data-efficient framework for studying thermophysical behavior in polymers and provide a foundation for future applications to broader polymer chemistries and properties.

\section{Data Availability}

\begin{suppinfo}
The supplementary information contains three tables having information (1) Monomer units, polymer chains, number of atoms, corresponding glass transition temperature ($T_g$), and coefficient of thermal expansion of various PE structures (SI-Table 1), (2) Details of monomer units, number of atoms in the training unit cell, Number of configurations sampled during MLFF training, training temperatures, and simulated larger structure used for T$_g$ calculations of the polymers (SI-Table 2) studied in this work, (3) Replica and two different cooling rate $T_g$ simulations for PE, (4) Weighted Flory–Fox extrapolation of $T_g$ to infinite molecular weight for PE, (5) Step-by-step MLFF generation protocol and extracting glass transition temperature, and 
(6) OTF-MLFF estimated coefficient of thermal expansion values below and above glass transition (SI-Table 3). Supplementary information also includes a description of error propagation estimation and a density-temperature plot for the simulated polymers (SI-Figure 3).
\end{suppinfo}

\section*{\large{AUTHOR INFORMATION}}
\textbf{Corresponding Author}\\
Rampi Ramprasad - \textit{School of Materials Science and Engineering, Georgia Institute of Technology, Atlanta, Georgia 30332, United States;} $^\ast$Email:{~ramprasad@gatech.edu}\\
\textbf{Authors}\\
Ashutosh Srivastava - \textit{School of Materials Science and Engineering, Georgia Institute of Technology, Atlanta, Georgia 30332, United States;} \\
Sakshi Agarwal - \textit{School of Materials Science and Engineering, Georgia Institute of Technology, Atlanta, Georgia 30332, United States;} \\
Shivank Shukla - \textit{School of Materials Science and Engineering, Georgia Institute of Technology, Atlanta, Georgia 30332, United States;} \\
Harikrishna Sahu - \textit{School of Materials Science and Engineering, Georgia Institute of Technology, Atlanta, Georgia 30332, United States;} 

\section{Author Contributions}
A.S. designed and established the overall simulation protocol, performed the DFT and OTF-MLFF calculations, carried out data analysis, and prepared the initial manuscript draft. S.A. contributed to the data analysis and assisted in preparing the original draft. S.S. performed the classical force-field simulations. H.S. supported the generation and preparation of polymer structures. R.R. conceived and supervised the project, provided overall guidance, and contributed to the interpretation of results.

\section{Acknowledgements}

This work is financially supported by the Office of Naval Research through a multidisciplinary university research initiative (MURI) grant N00014-20-1-2586 and the National Science Foundation through an NSF-SPEED grant 2515411. Computations were performed at Expanse (San Diego Supercomputing Center) through an allocation (DMR080044) from the Advanced Cyberinfrastructure Coordination Ecosystem: Services \& Support (ACCESS) program.
Discussions with Ayush Jain are also gratefully acknowledged.

\section{TOC Image}

\begin{figure}
  \centering
  \includegraphics[width=0.7\linewidth]{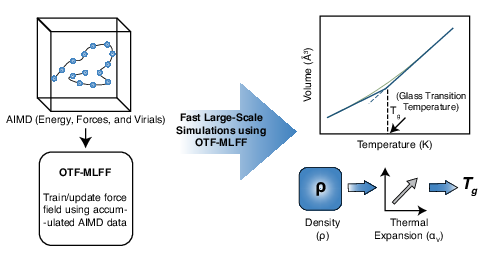} 
  \label{fig:TOC}
\end{figure}

\FloatBarrier


\providecommand{\latin}[1]{#1}
\makeatletter
\providecommand{\doi}
  {\begingroup\let\do\@makeother\dospecials
  \catcode`\{=1 \catcode`\}=2 \doi@aux}
\providecommand{\doi@aux}[1]{\endgroup\texttt{#1}}
\makeatother
\providecommand*\mcitethebibliography{\thebibliography}
\csname @ifundefined\endcsname{endmcitethebibliography}  {\let\endmcitethebibliography\endthebibliography}{}

\end{document}